\documentclass[lettersize,journal]{IEEEtran}

\usepackage[T1]{fontenc}
\usepackage{mathtools}
\usepackage{xurl}
\usepackage{amsmath}
\usepackage{acronym}  
\usepackage{amssymb}
\usepackage{caption}
\usepackage{amsthm}
\usepackage{amsfonts}
\usepackage{braket}
\DeclareMathOperator*{\argmin}{arg\,min}

\usepackage{upgreek}
\usepackage{dirtytalk}

\usepackage{optidef}
\usepackage{cite}
\usepackage{hyperref}
\usepackage{url}
\usepackage{xcolor}
\usepackage[nomain,acronym,shortcuts]{glossaries}\makeglossaries
\newcommand*{\acro}[3][]{\newacronym[#1]{#2}{#2}{#3}}
\glsdisablehyper
\acro{D2D}{device-to-device}
\acro{SIR}{signal-to-interference-ratio}
\acro{SINR}{signal-to-interference-plus-noise-ratio}
\acro{PCP}{Poisson cluster process}
\acro{CoMP}{coordinated multi-point}
\acro{BS}{base station} 
\acro{MD-CoMP}{macrodiversity CoMP transmission}
\acro{MAC}{medium-access-control}
\acro{JT-CoMP}{joint transmission CoMP}
\acro{CoMP-JT}{coordinated multipoint joint transmission}
\acro{SBS}{small base station}
\acro{MDSD}{multiple devices to a single device}
\acro{MDS}{maximum distance separable}
\acro{SCN}{small cell network}
\acro{PPP}{Poisson point process}
\acro{TCP}{Thomas cluster process}
\acro{CSI}{channel state information}
\acro{PDF}{probability distribution function}
\acro{PMF}{probability mass function}
\acro{RV}{random variable}
\acro{i.i.d.}{independently and identically distributed}
\acro{MBMS}{multimedia broadcasting multicasting service}
\acro{EE}{energy efficiency}
\acro{HCP}{hard-core placement}
\acro{CCDF}{complementary cumulative distribution function}
\acro{CDF}{cumulative distribution function}
\acro{PC}{probabilistic caching}
\acro{RC}{random caching}
\acro{CPF}{caching popular files} 
\acro{PGFL}{probability generating functional}
\acro{KKT}{Karush-Kuhn-Tucker}
\acro{PGF}{point generating function}
\acro{SCA}{successive convex approximation}
\acro{HD}{high-definition}
\acro{FHD}{full-high-definition}
\acro{UHD}{ultra-high-definition}
\acro{VR}{virtual reality}
\acro{AR}{augmented reality}
\acro{5G}{fifth-generation}
\acro{QoS}{quality-of-service}
\acro{QoE}{quality-of-experience}
\acro{IoT}{internet of things}
\acro{MHCPP}{Matern hardcore point process}
\acro{LoS}{line-of-sight}
\acro{NLoS}{non-line-of-sight}
\acro{PSD}{power spectral density}
\acro{MEC}{mobile edge computing}
\acro{E2E}{end-to-end}
\acro{THz}{terahertz}
\acro{CLT}{central limit theorem}
\acro{HQ}{High Quality}
\acro{eMBB}{enhanced mobile broadband}
\acro{URLLC}{ultra reliable low latency communications}
\acro{mmWave}{millimeter wave}
\acro{EVT}{extreme value theory}
\acro{GEV}{generalized extreme value}
\acro{HR2LLC}{high-rate and high-reliability low latency communications}
\acro{HF}{high frequency}
\acro{TVaR}{tail-value-at-risk}
\acro{UE}{user equipment}
\acro{RIS}{reconfigurable intelligent surfaces}
\acro{MIMO}{multiple-input and multiple-output}
\acro{HMD}{head-mounted display}
\acro{AI}{artificial intelligence}
\acro{ML}{machine learning}
\acro{HITRAN}{high resolution transmission}
\acro{OFDM}{orthogonal frequency division multiplexing}
\acro{PAPR}{peak-to-average power ratio}
\acro{MR}{mixed reality}
\acro{EVaR}{entropic value-at-risk}
\acro{DNN}{deep neural network}
\acro{MDP}{Markov decision process}
\acro{RL}{reinforcement learning}
\acro{RNN}{recurrent neural network}
\acro{ANN}{artificial neural networks}
\acro{LSTM}{long short-term memory}
\acro{ReLu}{rectified linear unit}
\acro{VaR}{value-at-risk}
\acro{SNR}{signal-to-noise ratio}
\acro{RF}{radio frequency}
\acro{CVaR}{conditional value-at-risk}
\acro{AVaR}{average value-at-risk}
\acro{ES}{expected shortfall}
\acro{AoI}{age of information}
\acro{PAoI}{peak age of information}
\acro{QoPE}{quality of physical experience}
\acro{LCFS}{last come first served}
\acro{FCFS}{first come first served}
\acro{SC-FDM}{single carrier frequency-division multiplexing}
\acro{ToA}{time of arrival}
\acro{MUSIC}{multiple signal classification}
\acro{IoE}{Internet of Everything}
\acro{6DoF}{six degrees of freedom}
\acro{OFDMA}{orthogonal frequency-division multiple access}
\acro{AoSA}{array of subarray}
\acro{XR}{extended reality}
\acro{AoA}{angle of arrival}
\acro{ULA}{uniform linear array}
\acro{AoD}{angle of departure}
\acro{EM}{electromagnetic}
\acro{IoI}{Internet of Intelligence}
\acro{MISO}{multiple-input and single-output}
\acro{CRAS}{connected robotics and autonomous systems}
\acro{FL}{federated learning}
\acro{GAN}{generative adversarial network}
\acro{mMTC}{massive machine type communications}
\acro{Tbps}{terabit per second}
\acro{NTN}{Non-Terrestrial Network}
\acro{SLAM}{simultaneous localization and mapping}
\acro{IS}{information shower}
\acro{OAM}{orbital angular momentum}
\acro{NOMA}{non-orthogonal mutliple access}
\acro{CPS}{cyber physical system}
\acro{IoNT}{Internet of Nano-Things}
\acro{TDD}{time division duplex}
\acro{CDMA}{code division multiple access}
\acro{OMA}{orthogonal multiple access}
\acro{VLC}{visible light communications}
\acro{UAV}{unmanned aerial vehicle}
\acro{LEO}{low earth orbit}
\acro{DL}{deep learning}
\acro{QSC}{quantum semantic communications}
\acro{QRAM}{quantum random access memory}
\acro{QC}{quantum computing}
\acro{QML}{quantum machine learning}
\acro{QIT}{quantum information theory}
\acro{QIP}{quantum information processing}
\acro{QCNs}{Quantum communication networks}
\acro{QCN}{quantum communication network}
\acro{QSE}{quantum semantic error}
\acro{QI}{quantum Internet}
\acro{NISQ}{noisy intermediate-scale quantum}

\usepackage{breqn}
\newcommand{\tr}{\mbox{Tr}}

\usepackage{enumitem}
\setlist{  
  listparindent=\parindent,
  parsep=0pt,
}
\DeclarePairedDelimiter\abs{\lvert}{\rvert}%
\newcommand{\norm}[1]{\left\lVert#1\right\rVert}

\usepackage{setspace}
\newcommand{\bmx}{{\boldsymbol x}}

\newcommand{\bms}{{\boldsymbol s}}
\newtheorem{lemma}{Lemma}

\begin{document}
\title{Quantum Semantic Communications for Resource-Efficient Quantum Networking}

\author{\IEEEauthorblockN{Mahdi Chehimi, Christina Chaccour, Christo Kurisummoottil Thomas, and Walid Saad}
\IEEEauthorblockA{Wireless@VT, Bradley Department of Electrical and Computer Engineering, Virginia Tech, Arlington, VA USA,\\
Emails: \{mahdic, christinac, walids\}@vt.edu}\vspace{-1.25cm}}
\vspace{-15mm}
\author{Mahdi Chehimi, Christina Chaccour, Christo Kurisummoottil Thomas, and Walid Saad
\thanks{M. Chehimi, C. K. Thomas, and W. Saad are with Bradley Department of Electrical and Computer Engineering, Virginia Tech, Arlington, VA 22203 USA. W. Saad is also with the Department of Computer Science and Engineering, Kyung Hee University, Yongin-si, Gyeonggido 17104, Rep. of Korea. (Emails: \{mahdic, christokt, walids\}@vt.edu).}
\thanks{C. Chaccour is with Ericsson, Inc., Plano, Texas, USA, Email: \protect{christina.chaccour@ericsson.com}.}
\vspace{-1cm}}

\maketitle\vspace{-0.4cm}
\begin{abstract}
Quantum communication networks (QCNs) utilize quantum mechanics for secure information transmission, but the reliance on fragile and expensive photonic quantum resources renders QCN resource optimization challenging. Unlike prior QCN works that relied on blindly compressing direct quantum embeddings of classical data, this letter proposes a novel \emph{quantum semantic communications (QSC)} framework exploiting advancements in quantum machine learning and quantum semantic representations to extracts and embed only the relevant information from classical data into minimal high-dimensional quantum states that are accurately communicated over quantum channels with quantum communication and semantic fidelity measures. Simulation results indicate that, compared to semantic-agnostic QCN schemes, the proposed framework achieves approximately 50-75\% reduction in quantum communication resources needed, while achieving a higher quantum semantic fidelity.
\end{abstract}\vspace{-1.5mm}
\begin{IEEEkeywords}
 \small Quantum networks, semantic communications, q-means clustering, semantic representation, fidelity.
\end{IEEEkeywords}
\IEEEpeerreviewmaketitle
\vspace{-0.3cm}
\section{Introduction}\vspace{-1mm}
Quantum communication networks (QCNs) utilize quantum mechanics principles to enhance information transfer. QCNs transmit data using quantum states that are entangled and can exist in a superposition of multiple states simultaneously, offering greater efficiency than classical networks \cite{chehimi2022physics}. However, these quantum states are fragile, expensive to produce, and vulnerable to environmental interference and loss, posing significant challenges in resource optimization for QCNs \cite{cacciapuoti2020entanglement}.

The majority of QCN models optimize the quantum resource allocation and network overall performance by embedding classical data into quantum states that are shared over quantum channels between distant nodes \cite{chandra2021direct,chehimi2021entanglement_rate_optimization,chehimi2023matching,chehimi2023scaling}. Additionally, numerous approaches have been proposed to develop resource-efficient QCNs, including strategies related to the physical layer and the quantum hardware, as well as techniques within the network control plane \cite{fan2021efficient,rozema2014quantum,yu2019quantum}. Here, quantum data compression has emerged as a preferred approach to minimize the usage of quantum resources, whereby an ensemble of quantum states is compressed, using methods like principal component analysis, into a smaller, potentially lower-dimensional, set that is communicated over QCNs \cite{fan2021efficient,rozema2014quantum,yu2019quantum}. However, existing approaches \cite{chandra2021direct,chehimi2021entanglement_rate_optimization,chehimi2023matching,chehimi2023scaling,fan2021efficient,rozema2014quantum,yu2019quantum} face several limitations as they rely on sending direct quantum embeddings of classical data in Hilbert spaces, which could include redundant or irrelevant embedded information. Furthermore, these methods do not consider the contextual and semantic meanings of the data; they simply transmit the raw data or its compressed form without understanding its content. Here, we observe that \emph{no prior work has utilized the principles of semantic communications in QCNs to make them more resource-efficient.}

\begin{figure*}\centering
\includegraphics[width=\textwidth]{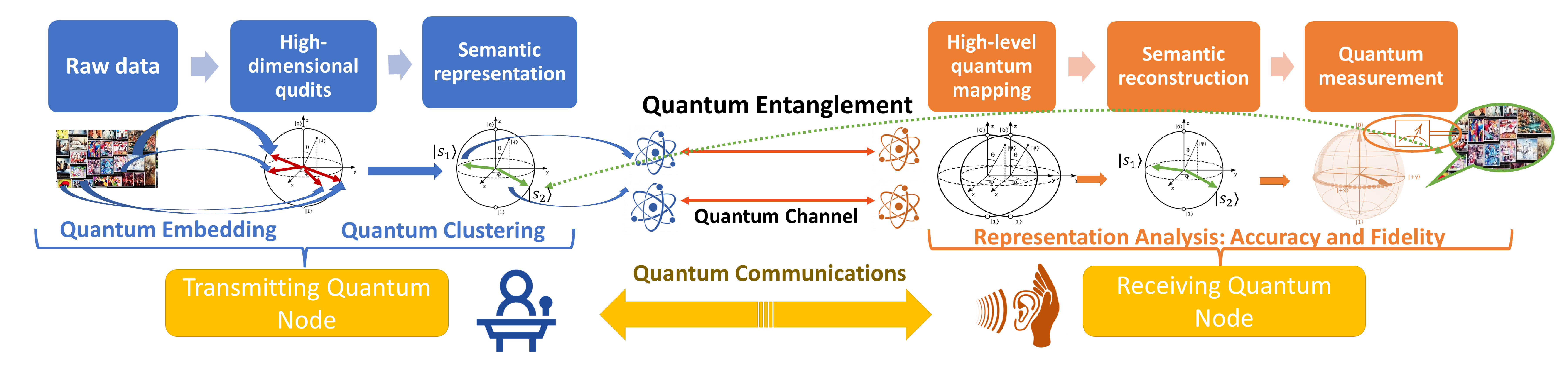}
\captionsetup{justification=centering,margin=2cm}
\vskip -0.15in\caption{Illustrative figure showcasing the proposed QSC framework.}
\label{fig_areas}
\vskip -0.25in
\end{figure*}

Towards this goal, the main contribution of this letter is a novel resource-efficient QCN framework, dubbed the \emph{\ac{QSC}} framework. This framework draws upon recent advancements in two key quantum information science areas. First, it utilizes high-dimensional quantum information and \ac{QML} to extract underlying structures of classical data, capitalizing on its efficiency in identifying atypical data patterns and surpassing classical machine learning speeds \cite{biamonte2017quantum}. Second, the QSC framework delves into quantum semantic representations, highlighting quantum mechanics' fundamental role in vector modeling and linear algebraic semantics \cite{widdows2021quantum,tetlow2022towards}. Thus, rather than merely transmitting raw data in compressed quantum states, the QSC framework intelligently extracts semantic information from the data. It then transmits only the essential quantum semantic representations over quantum channels, thereby resulting in more resource-efficient QCNs while maintaining high accuracy. Specifically, we provide a systematic approach for assessing and optimizing the \emph{minimality} of quantum communication resources needed (e.g., entangled quantum states), and the \emph{accuracy} of those resources in terms of quantum communication and semantic fidelity, showcasing the tradeoffs that exist. Simulation results validates that the QSC framework results in minimal quantum communication resources, saving 50-75\% of the resources compared to semantic-agnostic QCNs, while achieving higher quantum semantic fidelity. The proposed framework provides a promising direction for researchers and engineers to explore the potential of QML and \ac{QIT} in reducing the resources required in QCNs. Figure \ref{fig_areas} illustrates the proposed QSC framework, which starts at the transmitter, where raw data is embedded into qudits in high-dimensional Hilbert spaces. Subsequent quantum clustering extracts useful semantic concepts, which are then mapped into efficient semantic representations for transmission via entangled qudits over quantum channels. At the receiver end, the fidelity and accuracy of the quantum mapping are verified, followed by the reconstruction of semantic representations and derivation of semantic concepts through quantum measurements.

\vspace{-0.3cm}
\section{System Model}\label{sec_system_model}\vspace{-1.5mm}
Consider a QCN in which a transmitting quantum node, $S$, has large amounts of raw classical data, in dataset $\mathcal{X}$ of images, text, etc., with $\abs{\mathcal{X}}$ samples each of dimension $N$, which it embeds into quantum states. The transmitting node's intent is to construct a semantic representation of its quantum states and to convey it to a remote receiving quantum node, $R$. 
In contrast to conventional QCN frameworks, where the receiver typically aims to execute quantum gates and measurements for the precise reconstruction of the embedded data structure within quantum states, our framework distinguishes itself. Here, the receiver's goal is to draw specific logical conclusions \cite{ChristoTWCArxiv2022}, which may not necessarily involve reconstructing the entire data samples. Therefore, the primary objective of the quantum receiving node is to \emph{comprehend the underlying semantic information conveyed in quantum states}~\cite{chaccour2022less}, facilitating the execution of logical conclusions at its end. To develop the notion of \emph{quantum semantics}, we mainly leverage the characteristics of high-dimensional quantum states in representing semantics jointly with the use of \ac{QML} techniques. Next, we delve into our novel quantum semantic representation methodology, which relies on the principles of \ac{QIT}, high-dimensional Hilbert spaces, and \ac{QML}.

\vspace{-0.5cm}
\subsection{Quantum Data Embedding}\label{sec_embeddings}\vspace{-0.1cm}
In classical communications, multiplexing enables combining multiple messages into a shared medium (frequency, time, or space). Similarly, in the quantum world, a quantum state that can hold a superposition of $d_1$ basis vectors is called a \emph{qudit}. 
This superposition inherently enables us to enhance the capacity of the information contained within a quantum state. The general representation of a qudit is a vector $\ket{\varphi}$ in a $d_1$-dimensional Hilbert space $\mathcal{H}_{d_1}$ represented as: 
\begin{equation}\small
    \ket{\varphi} = a_0 \ket{0} + a_1 \ket{1} + a_2 \ket{2} + ... + a_{d_1-1} \ket{d_1-1},
\end{equation}
where $|{a_0}|^2 + |{a_1}|^2 + ... + |{a_{d_1-1}}|^2 = 1$. The quantum state \(\ket{\varphi} \) can also be represented within a density matrix as a sum of outer products of basis states, i.e., \( \rho \equiv \sum_{i=0}^{d_1-1} p_i \ket{i}\bra{i} \), where \( p_i \) represents the probability of the state being in the \( \ket{i} \) state. The von Neumann entropy can be defined as \cite{tetlow2022towards}, $\mathbb{H}_q = -\tr\left(\rho\ln \rho\right)$, where ``$\tr$" denotes the trace of a matrix. After pre-processing our classical data by applying contrastive learning to disentangle semantically-rich data points from semantically-poor data points \cite{chaccour2022disentangling}, we then encode the classical data $\bmx \in \mathcal{X}$ into quantum states $\ket{\psi(\bmx)} = \ket{y}$ using a quantum feature map $\psi: \mathcal{X} \rightarrow \mathcal{Y}$, where $\mathcal{Y} \subset \mathcal{H}_{d_1}$. Here, the Hilbert space's dimension, $d_1$, is much larger than $N$, the dimension of the data in $\mathcal{X}$. This is due to the potential lack of linear separability in the raw data space $\mathcal{X}$ for the data points $\bmx$. Consequently, transforming it into a higher-dimensional space $\mathcal{Y}$ becomes essential, facilitating the clustering of data points based on the semantic concepts, as described later. The quantum feature map $\psi$, which maps $\bmx\rightarrow \ket{y}$, is implemented via a quantum circuit $U_{\psi,d_1}(\bmx)$, called the quantum-embedding circuit. This circuit feeds the classical data $\bmx$ to quantum gates applied on ground states, $\ket{0_{d_1}}$, in $\mathcal{H}_{d_1}$. To ensure computational feasibility and to avoid exponentially growing Hilbert spaces, quantum embedding circuits in the QSC framework are designed such that they compromise between the number of encoding qudits, and the dimensionality, $d_1$, of the Hilbert space. This optimization is mainly dependent on the size and dimensionality of the classical data, along with the available quantum computing powers. Essentially, such a quantum feature map yields the quantum embedded states $\ket{y}$ according to: $U_{\psi,d_1}(\bmx)\ket{0_{d_1}} = \ket{y}$ \cite{schuld2019quantum}. Subsequently, these quantum states are stored in tailored \ac{QRAM} structures. After embedding the data into quantum states, the next step is apply QML to construct efficient quantum semantic representations, as described next.


\vspace{-0.45cm}
\subsection{Quantum Clustering for Semantics' Extraction}\label{sec_clustering}\vspace{-0.05cm}
Existing semantic-agnostic QCN frameworks \cite{chandra2021direct,chehimi2021entanglement_rate_optimization,chehimi2023matching,chehimi2023scaling,fan2021efficient,rozema2014quantum}, \say{blindly} compress classical datasets into quantum states, without scrutinizing their underlying structure. In contrast, our proposed \ac{QSC} framework performs the following steps to extract and represent the semantic information in a minimalist approach that improves the overall QCN resource-efficiency.

First, the quantum $k$-means clustering algorithm \cite{kerenidis2019q} is performed on the \ac{QRAM}. This algorithm identifies data samples that share various similarity in their contextual meaning, and assigns them to unique clusters ($K$ of them) with centroid vectors $\ket{\phi_i}$, where $i\in\{1,2,...,K\}$ capturing those meanings or semantics in a high-dimensional representation of semantic concepts. To optimize the representation of those semantic concepts, we formulate a modified objective function for the quantum $k-$means clustering algorithm that represents a compromise between reducing the number of semantic concepts identified and clustering the data points $\ket{y} \in \mathcal{Y}$. We denote any cluster that includes the centroid along with a subset of the quantum states $\ket{y}$ as $c_k$.

\subsubsection{Problem formulation for quantum k-means clustering}
The probability of a data point belonging to one of the $K$ clusters, e.g., cluster $c_k$, is defined as $p(c_k)$, which depends on the data distribution $p(\bmx)$. Further, we define the cluster entropy $\mathbb{H}(c) = -\sum\limits_k p(c_k)\ln p(c_k)$ as the Shannon entropy, since $c$ represents the clusters or regions, not quantum states. Moreover, we define the probability that a data point $\ket{y_i}$ belongs to the cluster $c_k$ as $p(c_k\mid\ket{y_i})$. We modify the objective of the $k$-means clustering algorithm as follows:
\begin{equation}\footnotesize
\begin{aligned}
J(p(c),p(c\mid\ket{y}),\beta) &= \sum\limits_{i} \sum\limits_{k} p(c_k\mid\ket{y_i})\norm{\ket{y_i}-\ket{\phi_i}}^2 \\ &- \beta\left(\underbrace{\mathbb{H}(c) - \mathbb{H}(c \mid \ket{y})}_{\mathbb{I}(\ket{y};c)}\right), \label{eq_kmeans}
\end{aligned}
\end{equation}
where $ \mathbb{H}(c \mid \ket{y}) = -\sum\limits_{i}\sum\limits_{k}p(c_k\mid\ket{y_i})\ln p(c_k\mid\ket{y_i})$. The penalty term $\mathbb{I}(\ket{y};c)$ ensures that the semantic information conveyed by the $k$-means clustering that identifies the semantic concepts is maximized.
\subsubsection{Solution to \eqref{eq_kmeans}}
Here, the stored quantum vectors are initialized to different clusters either in an arbitrary fashion or by utilizing efficient heuristic approaches. Then, multiple iterations are performed such that in each iteration, the goal is to minimize the loss function in \eqref{eq_kmeans}, which ensures that each vector is assigned to the cluster with the closest centroid. This process ensures that semantically similar data vectors are clustered together (as guaranteed by the penalty term in \eqref{eq_kmeans}). This clustering aids in uncovering hidden patterns that might remain undiscovered in the raw classical data without the utilization of \ac{QML}.
All the quantum states belonging to the same cluster form a semantic space and are semantically similar, as each centroid represents a semantic concept. Henceforth, the set $\Phi$ of $\ket{\phi_i}, \forall i$ forms a common language, composed of the \emph{syntactic space} \cite{ChristoTWCArxiv2022}. Along the lines of \cite{ChristoTWCArxiv2022}, the amount of semantic information conveyed by any cluster $c_k$ can be written as the average across of the information represented by all the elements in the cluster $c_i$: $   \mathbb{S}(c_k) = \sum\limits_{\ket{y}\in \mathcal{Y}}  p(\ket{y})\log\frac{p(\ket{y}\mid c_k)}{p(\ket{y})}.$
Given that the number of semantically distinct concepts $K \leq \abs{\mathcal{X}}$, we chose to encode the semantic concepts into semantic representations $\ket{s} \in \mathcal{S}$ that exists in a lower dimensional Hilbert space for quantum communication $\mathcal{H}_{d_2}$, where $d_2\leq d_1$. Semantic representation ensures that all the concepts that are identical from the receiver's perspective are encoded with the same representation that gets communicated over the QCN, thus saving significant amount of quantum communication resources. A semantic representation is defined as the mapping, $f: \mathcal{Y} \,\cup\, \Phi \rightarrow \mathcal{S}$ and is implemented as a probabilistic mapping $p(\ket{s_i}\mid c_k)$. The number of \emph{quantum communication resources} $C = \abs{\mathcal{S}}$ corresponds to the number of quantum states transmitted, which can be less than the number of clusters $K$ as the semantic representation mapping $f$ consolidates all semantically similar concepts and represents them using the same $\ket{s_i} \in \mathcal{S}$.
\vspace{-2mm}\begin{lemma}
\label{theorem_se_rep}
For a quantum state space $\mathcal{Y}$, and a semantic context distribution $p(c)$, the average number of quantum communication resources, $C$, required to
represent the state description in the QSC framework can be
bounded as: \begin{equation}\vspace{-1mm}
\sum\limits_{c_k\in \mathcal{C}}p(c_k) \mathbb{H}_q(\ket{y}\mid c_k) \leq 
C \leq \ln d_2
,\label{eq_RepLength_Theorem}
 \vspace{-0mm} \end{equation} while for a semantic-agnostic QCN with amplitude encoding, the bounds are: \vspace{-2mm}\begin{equation}
\mathbb{H}_q(\ket{y}) \leq C \leq  \ln d_1.
\vspace{-1mm}\label{eq_classicalbound}\end{equation}
 Comparing \eqref{eq_RepLength_Theorem} and \eqref{eq_classicalbound}, the bounds for the QSC framework are lower compared to that of semantic-agnostic QCN systems that do not extract the data semantics.
 \vspace{-0mm} 
  \vspace{-0mm}\begin{IEEEproof}
    \vspace{-0mm} See Appendix~\ref{appendix_proof_theorem_se_rep}.
   \vspace{-0mm}\end{IEEEproof}
\end{lemma}\vspace{-0.25cm}


Using the semantic information measures in \cite{ChristoTWCArxiv2022}, we write the average semantic information conveyed by $\ket{s_j}$ as:
\begin{equation}\small
    \mathbb{E}_{\bms} \mathbb{S}(c_k;\ket{s_j}) = \sum\limits_{c_k}  p(c_k)\mathbb{S}(c_k)p(\ket{s_j})\log \frac{p(\ket{s_j}\mid c_k)}{p(\ket{s_j})}.
\end{equation}

\vspace{-0.55cm}
\subsection{Quantum Communication of Semantic Representations}
As shown in Fig.~\ref{fig_areas}, the constructed quantum semantic representations in the form of $d_2$-dimensional quantum states must be transmitted through quantum channels to the receiving quantum node. The quantum communication process must preserve the accuracy of the quantum semantic features during the transmission and reception process. 

To deploy high-dimensional quantum states, qudits, using photons of light, we leverage the concept of \ac{OAM}. \ac{OAM} is a physical property of an electromagnetic wave and it corresponds to the phase of its angular momentum. The topological charges of \ac{OAM}, i.e., the \emph{modes} of \ac{OAM}, are orthogonal and enable us to exploit $d$ orthogonal basis vectors that represent a qudit in a $d$-dimensional Hilbert space $\mathcal{H}_d$. In other words, an arbitrary $l$-dimensional qudit is generated via $l$ \ac{OAM} modes. In fact, the quantum number $l$, which represents a topological charge in \ac{OAM}, is unbounded, and can thus yield arbitrarily large Hilbert spaces. Nonetheless, due to some experimental considerations, some constraints may be imposed on $l$ \cite{cozzolino2019high}.\footnote{\scriptsize Diverse methods for implementing qudits extend beyond \ac{OAM} encoding, encompassing techniques like spontaneous parametric down-conversion and waveshaping devices. For hardware-specific details, see \cite{cozzolino2019high} and references therein.} Thus, pairs of entangled photons with opposite \ac{OAM} quantum numbers $l_1 = -l_2$ are generated on the transmitter's side. The theoretical states produced for those photons are given by $\ket{\Psi} = \sum_{l = -d/2}^{l=d/2}{a_l\ket{l_1}\ket{-l_2}},$ where $\ket{\pm l_i}$ represents the states of the two generated photons with \ac{OAM} $\pm l$ and the complex probability amplitudes are represented by $a_l$.

To initiate a quantum entanglement link between the transmitter and receiver, the transmitter generates an entangled pair of photons using OAM, and one of those photons is transmitted over a quantum channel (fiber or free-space optical channel) to the quantum receiving node that stores it in a quantum memory. Given that the entangled link is now established, the transmitter maps each of the $\abs{S}$ semantic-representing $d_2$-dimensional quantum state vectors to one of its entangled respective photons through quantum swapping. 
Finally, the quantum teleportation protocol is applied to transfer the semantics to the receiver, and both end nodes perform quantum measurements and apply quantum gates to reconstruct the embedded semantics and recover, the context of the raw data.
The receiver extracts the semantic concepts using the mapping, $g:\mathcal{S} \rightarrow \mathcal{Y}\cup{\boldsymbol \Phi}$, which is represented using the the probabilistic mapping $p(c_k\mid \ket{s_i})$. Next, we discuss our proposed novel approaches to assess the performance of the \ac{QSC} framework that involves the design of semantic encoding function $f$ and decoding function $g$.
\vspace{-0.35cm}
\section{QSC Problem Formulation and Performance Assessment Framework}\label{sec_performance_assessment}\vspace{-1mm}
As discussed earlier, the QSC framework ensures minimality of quantum communication resources by extracting and compressing the semantic representations of the data, unlike existing semantic-agnostic QCNs. Moreover, to assess the accuracy of the QSC performance within the quantum semantics' extraction, transmission, reception, and decoding processes. To do so, we must the error sources encountered during both the quantum communication errors and the quantum semantic errors present in practical setups in the QSC framework. 

First, regarding the quantum communication aspect, today's quantum computers are \ac{NISQ} devices that incorporate noise and various unavoidable losses that affect the quantum circuits employed in quantum embedding, entanglement generation, and entanglement measurement operations. A worst-case general quantum channel model that captures such noise sources in QCNs is the \emph{quantum depolarizing noise} model, where a quantum state $\rho$ is preserved with a probability $1-\lambda$ and lost with probability \(\lambda\), where $\lambda$ is a real-valued parameter \cite{king2003capacity}. This model, denoted as \(\Delta_\lambda\), maps a quantum state $\rho\in\mathcal{H}_d$ into a linear combination that comprises $\rho$ itself, and the $d\times d$ identity matrix $\boldsymbol{I}_d$. It is a trace-preserving, completely positive map, given by: $ \Delta_\lambda(\rho) = (1-\lambda)\rho + \frac{\lambda}{d}\boldsymbol{I}_d.$
It is important to note that parameter $\lambda$ is bounded by: $-\frac{1}{d^2-1}\leq\lambda\leq1,$ so as to guarantee the complete positivity condition. The quality of the communicated quantum resources is quantified using the \emph{quantum communication fidelity}, $F_c$, of the considered received quantum states which captures the unavoidable noise that every quantum state undergoes when transmitted over quantum channels. In this regard, when the generated entangled pairs of qudits are considered to have maximally entangled measurements and channels, the average quantum communication fidelity that corresponds for the depolarizing noise will be \cite{king2003capacity} $\langle F_c\rangle = 1 - \frac{d-1}{d}\lambda$, where $d$ is the dimension of the qudits. If a low fidelity is encountered, entanglement purification protocols may be applied \cite{pan2001entanglement} to achieve a desired minimum quality of the entanglement connection.
    Meanwhile, when the entanglement connection between the transmitter and receiver is characterized with a high fidelity, then the semantic-representing quantum states will be accurately received. 
    
    However, the quantum semantic extraction, representation, and decoding steps also encounter several errors, which affect the quality of the extracted semantics. To quantify such semantic errors, we define the quantum semantic fidelity as:
\begin{equation}\small
\langle F_s(\ket{y_i},\ket{y_j})\rangle = 1 - \frac{d_1-1}{d_1}p_s(\ket{y_i},\ket{y_j}),
\end{equation}
where $p_s$ denotes the probability of semantic noise (a measure of the semantic reconstruction quality) and is defined as $p_s=1-Z_{y_iy_j}$. The semantic similarity metric, $Z_{y_iy_j}$, between two quantum states, $\ket{y_i}$ and $\ket{y_j}$ in a cluster, can be defined as the overlap in terms of the semantic concepts that represent the embedded data points. Analytically, we define it as: 
\begin{equation}\small
\vspace{-1mm} \footnotesize Z_{y_iy_j} = \sqrt{1 - \sum\limits_kp(c_k)\abs{p\left(c_k\mid \ket{y_i}\right)-p\left(c_k\mid \ket{y_j}\right)}}\vspace{-0mm}
\end{equation}
The above definition is appropriate since it means that $p\left(c_k\mid \ket{y_i}\right) = p\left(c_k\mid \ket{y_j}\right)$, then $Z_{ij}=1$ and when the domain of $p\left(c_k\mid \ket{y_i}\right) $ and $p\left(c_k\mid \ket{y_j}\right)$ does not overlap, $Z_{ij} = 0$. Additionally, with an increased number of clusters, the overlap between the domains of $p\left(c_k\mid \ket{y_i}\right) $ and $p\left(c_k\mid \ket{y_j}\right)$ decreases. Consequently, the quantity $\sum\limits_kp(c_k)\abs{p\left(c_k\mid \ket{y_i}\right)-p\left(c_k\mid \ket{y_j}\right)}$ increases, leading to lower semantic similarity.

In general, there is a tradeoff between \emph{minimality} and \emph{accuracy} in the QSC framework as a smaller number of quantum communication resources leads to smaller quantum semantic fidelity. Thus, we formulate the QSC minimality-accuracy tradeoff problem as follows:
\begin{equation}\vspace{-1mm}\small
\begin{aligned}
    \left[f^{*}, g^{*}\right] =& \argmin\limits_{f,g} \mathbb{E}\left[\mathbb{S}(c;\ket{s})\right] \\
    \mbox{subject to}\,\, &\mathbb{E}\left[F_s(c_k,\widehat{c}_k)\right] \geq S_{\tau}, \\
&\mathbb{E}\left[F_c(\ket{s_j},\ket{\widehat{s}_j})\right] \geq D_{\tau}, \\
&\sum\limits_{\ket{s_i}}p(\ket{s_i}\mid c_k) = 1, \forall c_k,
\label{eq_min_semanticinfo}
    \end{aligned}\vspace{-2mm}
\end{equation}
where, $S_{\tau}$ and $D_{\tau}$ represent the minimum required fidelity measures for quantum semantics and quantum communication, respectively. The objective function in \eqref{eq_min_semanticinfo} seeks to minimize the communicated average semantic information $\mathbb{E}\left[\mathbb{S}(c;\ket{s})\right]$, while the constraints ensure that the quantum communication and semantic fidelity remains above a minimum threshold, sufficient for the receiver to reconstruct the semantic information.

To solve \eqref{eq_min_semanticinfo}, given a receiver policy $p(c_k\mid \ket{s_i})$, we optimize the transmit semantic encoder by maximizing the Lagrangian. We define ${\ket{\widehat{s}_i}}$ as the received quantum state.  
\vspace{-1mm}\begin{equation}\small
    \vspace{-0mm}\begin{aligned}
&p(\ket{s_i}\mid c_k)^* = \,\arg\min\limits_{p(\ket{s_i}\mid c_k)}\mathcal{L}(p(\ket{s_i}\mid c_k),p(c_k\mid \ket{s_i})),\\ &\mbox{where},\,\, \mathcal{L}(p(\ket{s_i}\mid c_k),p(c_k\mid \ket{s_i})) = \,\mathbb{E}\left[\mathbb{S}(c;\ket{s})\right] + \\ &\lambda_s p(c_k) F_s(c_k,\widehat{c}_k) + \lambda_c p(c_k)p(\ket{s_i}\mid c_k)F_c(\ket{s_i},\ket{\widehat{s}}_i)\\ &+ \sum\limits_{c_k} \alpha_{c_k}\left[\sum\limits_{\ket{s_i}}p(\ket{s_i}\mid c_k)\right].
\end{aligned}\vspace{-1mm}
\end{equation}
It can be validated that $\mathcal{L}(p(\ket{s_i}\mid c_k),p(c_k\mid \ket{s_i}))$ is concave in $p(\ket{s_i}\mid c_k)$
(follows from showing that term of the form $x \log \frac{x}{a}$ is
convex), hence the global maximum is obtained at the point
where derivative of $\mathcal{L}$ is zero. By taking the derivative and equating to zero, the transmitter encoder $f$ is given by:
\begin{equation}\label{eq_speakTransmit_sol}\small
    p_{\ket{s_j}\mid c_k}^* = p(\ket{s_j})\exp\left(\frac{-\left(F_{c_k,s_j}-\alpha_{c_k}\right)}{\mathbb{S}(c_k)p(c_k)}-1\right),
\end{equation}
where $F_{c_k,s_j} = \lambda_c (F_c(\ket{s_j},\ket{\widehat{s}_j}))+\lambda_s (F_s(c_k,\widehat{c}_k))$.

For the receiver side, we can show that optimizing the rate-distortion objective results in the parameterized solution as \eqref{eq_receiver_sol}, which is inspired from the results in \cite{StavrouTCOM2023}. 
\begin{equation} 
p_{c_k | \ket{s_j}}^* = \frac{e^{F_{c_k,s_j}} \nu^{*}(\widehat{c}_k, \ket{\widehat{s}_j})}
{\sum_{\widehat{c}_k, \ket{\widehat{s}_j}} \left[ e^{F_{c_k,s_j}} \nu^{*}(\widehat{c}_k, \ket{\widehat{s}_j}) \right]}
\label{eq_receiver_sol}
\end{equation}\vspace{-0.3cm}

The iterative optimization process between $p(\ket{s_j}\mid c_k)$ and $p(c_k\mid \ket{s_j})$ leads to a locally optimal solution, given that each sub-problem involved is convex.

\vspace{-0.45cm}\section{Simulation Results and Analysis}\label{sec_simulations}\vspace{-0.2cm}

\begin{figure}[t]
\begin{center}\vspace{-0.05in}
\centerline{\includegraphics[width=\columnwidth]{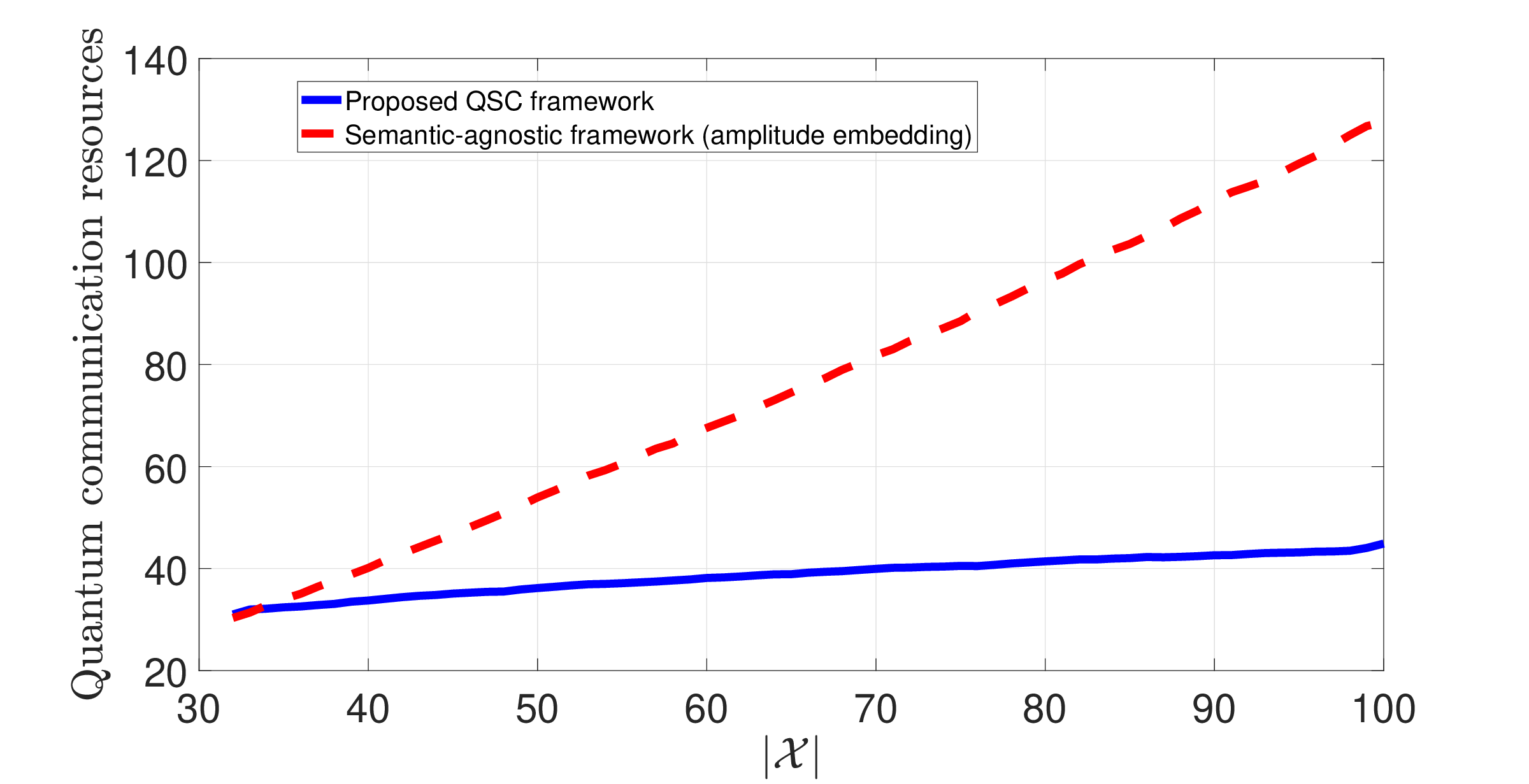}}\vspace{-0.1in}
\caption{\small{Comparison of communication resources for QSC and semantic agnostic networks (for which $d_2=3$).}}
\label{fig_comm_resources_QSC_vs_agnostic}
\end{center}\vspace{-0.38cm}
\end{figure}

For our simulations, we consider a scenario in which the data traffic is modeled according to 3GPP TSG-RAN$1\#48$ R1-070674 \cite{3gpp2007lte}. This leads to a complex data traffic model suitable for 5G wireless services. Here, we particularly model a gaming data traffic, where the data dimension is 3, and packet size varies according to the largest extreme value probability distribution $f_x = \frac{1}{b}e^{-\frac{x-\mu}{\sigma}}e^{-e^{-\frac{x-\mu}{\sigma}}},$ with a mean $\mu = 45$ bytes, and a standard deviation $\sigma = 5.7$ bytes \cite{navarro2020survey}. 

First, in Figure \ref{fig_comm_resources_QSC_vs_agnostic}, we compare the quantum communication resources needed for QSC and semantic-agnostic frameworks. We observe that as the data traffic increases (represented by $\abs{\mathcal{X}}$), the amount of semantic concepts extracted will increase which causes the monotonic increase of the quantum communication resources. 
Nevertheless, with the application of quantum $k$-means clustering, which extracts only the pertinent semantic concepts for communication, the required amount of communication resources is significantly reduced, for instance, by approximately $50\%$ at $\abs{\mathcal{X}} = 70$, and $75\%$ at $\abs{\mathcal{X}} = 100$ compared to semantic-agnostic QCNs. 
\vspace{0mm}\begin{figure}[t]
\begin{center}\vspace{-0.27in}
\centerline{\includegraphics[width=\columnwidth]{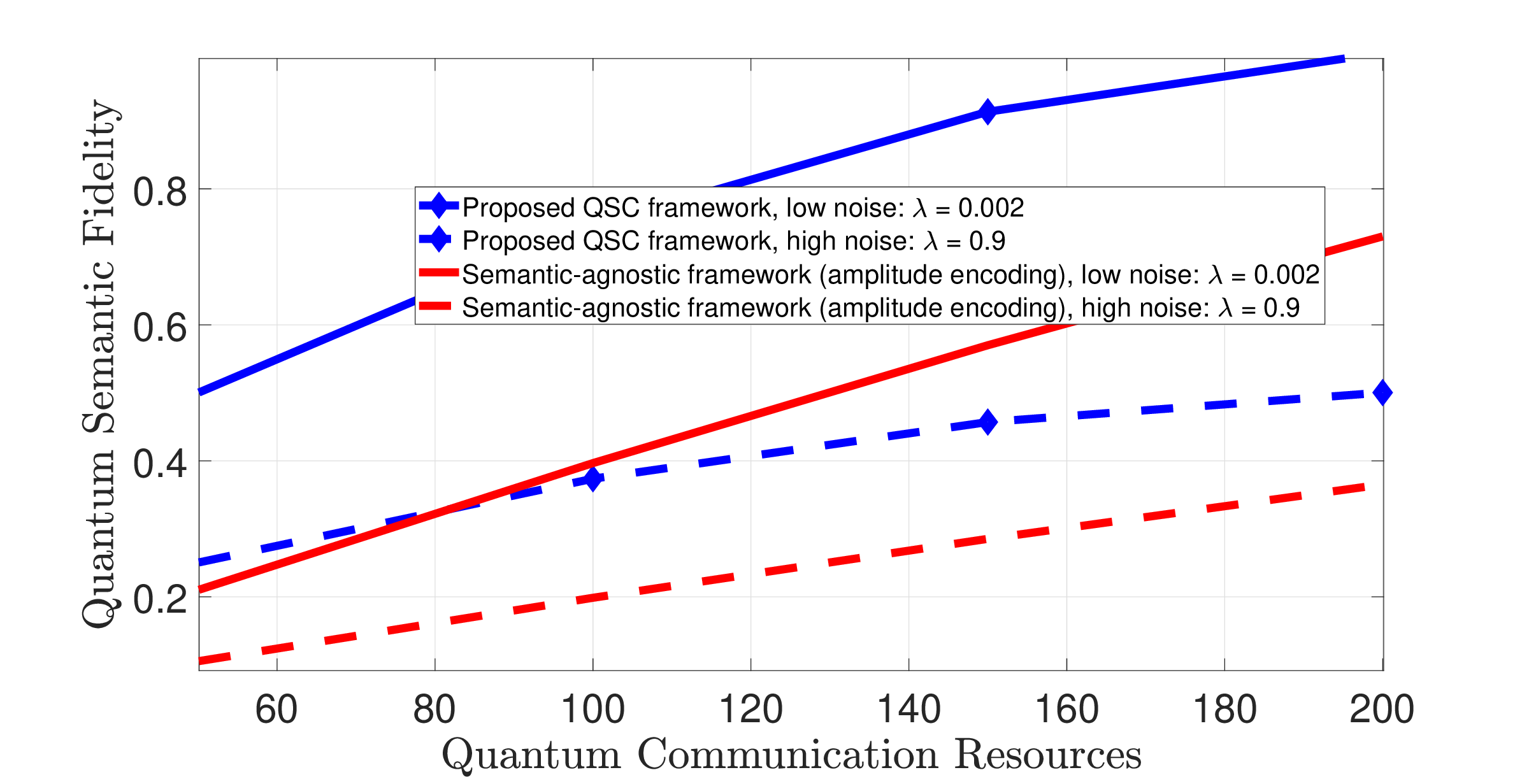}}\vspace{-0.1in}
\caption{\small{Semantic fidelity vs quantum communication resources used, for fixed $\abs{\mathcal{X}}$.}}
\label{fig_semanticfidelity}
\end{center}\vspace{-1.15cm}
\end{figure}

In Figure \ref{fig_semanticfidelity}, we show the quantum semantic fidelity achieved against the amount of quantum communication resources used for $\abs{\mathcal{X}} = 500$. At low noise, to achieve a quantum semantic fidelity of $0.7$, QSC requires around $50\%$ quantum communication resources compared to semantic-agnostic QCNs using pruning data compression without any semantic concept extraction. This demonstrate the advantages of QSC accurately sending and reconstructing semantic information.

\vspace{-0.5cm}\section{Conclusion}\label{sec_conclusion}\vspace{-2mm}
In this letter, we have proposed a novel \ac{QSC} framework for resource-efficient QCNs. The QSC framework utilizes QML and QIT to extract and minimally represent semantic information in data, then accurately share it over quantum channels. The QSC framework analyzes the minimality-accuracy tradeoff while capturing quantum communication and semantic fidelity. Simulation results show that the QSC framework results in minimal quantum communication resources, saving 50-75\% of the resources compared to semantic-agnostic QCNs, while achieving higher quantum semantic fidelity.

\vspace{-4mm}\begin{spacing}{0.7}
\bibliographystyle{IEEEtran}
\def\baselinestretch{0.60}
\bibliography{References}\vspace{-2mm}

\begin{thebibliography}{10}
\providecommand{\url}[1]{#1}
\csname url@samestyle\endcsname
\providecommand{\newblock}{\relax}
\providecommand{\bibinfo}[2]{#2}
\providecommand{\BIBentrySTDinterwordspacing}{\spaceskip=0pt\relax}
\providecommand{\BIBentryALTinterwordstretchfactor}{4}
\providecommand{\BIBentryALTinterwordspacing}{\spaceskip=\fontdimen2\font plus
\BIBentryALTinterwordstretchfactor\fontdimen3\font minus \fontdimen4\font\relax}
\providecommand{\BIBforeignlanguage}[2]{{%
\expandafter\ifx\csname l@#1\endcsname\relax
\typeout{** WARNING: IEEEtran.bst: No hyphenation pattern has been}%
\typeout{** loaded for the language `#1'. Using the pattern for}%
\typeout{** the default language instead.}%
\else
\language=\csname l@#1\endcsname
\fi
#2}}
\providecommand{\BIBdecl}{\relax}
\BIBdecl

\bibitem{chehimi2022physics}
M.~Chehimi and W.~Saad, ``Physics-informed quantum communication networks: A vision towards the quantum internet,'' \emph{IEEE network}, pp. 134--142, Sep. 2022.

\bibitem{cacciapuoti2020entanglement}
A.~S. Cacciapuoti \emph{et~al.}, ``When entanglement meets classical communications: Quantum teleportation for the quantum {Internet},'' \emph{IEEE Transactions on Communications}, vol.~68, no.~6, pp. 3808--3833, 2020.

\bibitem{chandra2021direct}
D.~Chandra \emph{et~al.}, ``Direct quantum communications in the presence of realistic noisy entanglement,'' \emph{IEEE Transactions on Communications}, vol.~70, no.~1, pp. 469--484, 2021.

\bibitem{chehimi2021entanglement_rate_optimization}
M.~Chehimi and W.~Saad, ``Entanglement rate optimization in heterogeneous quantum communication networks,'' in \emph{17th International Symposium on Wireless Communication Systems (ISWCS)}.\hskip 1em plus 0.5em minus 0.4em\relax IEEE, 2021, pp. 1--6.

\bibitem{chehimi2023matching}
M.~Chehimi, B.~Simon, W.~Saad, A.~Klein, D.~Towsley, and M.~Debbah, ``Matching game for optimized association in quantum communication networks,'' in \emph{Proc. of IEEE Global Communications Conference (Globecom)}, Kuala Lumpur, Malaysia, Dec. 2023.

\bibitem{chehimi2023scaling}
M.~Chehimi, S.~Pouryousef, N.~Panigrahy, D.~Towsley, and W.~Saad, ``Scaling limits of quantum repeater networks,'' in \emph{Proc. of IEEE International Conference on Quantum Computing and Engineering (QCE)}, Bellevue, WA USA, Sep. 2023.

\bibitem{fan2021efficient}
C.-R. Fan \emph{et~al.}, ``Efficient multi-qubit quantum data compression,'' \emph{Quantum Engineering}, vol.~3, no.~2, p. e67, 2021.

\bibitem{rozema2014quantum}
L.~A. Rozema, D.~H. Mahler, A.~Hayat, P.~S. Turner, and A.~M. Steinberg, ``Quantum data compression of a qubit ensemble,'' \emph{Physical review letters}, vol. 113, no.~16, p. 160504, 2014.

\bibitem{yu2019quantum}
C.-H. Yu \emph{et~al.}, ``Quantum data compression by principal component analysis,'' \emph{Quantum Information Processing}, vol.~18, pp. 1--20, 2019.

\bibitem{biamonte2017quantum}
J.~Biamonte \emph{et~al.}, ``Quantum machine learning,'' \emph{Nature}, vol. 549, no. 7671, pp. 195--202, Sep. 2017.

\bibitem{widdows2021quantum}
D.~Widdows \emph{et~al.}, ``Quantum mathematics in artificial intelligence,'' \emph{Journal of Artificial Intelligence Research}, vol.~72, pp. 1307--1341, Dec. 2021.

\bibitem{tetlow2022towards}
P.~Tetlow, D.~Garg, L.~Chase, M.~Mattingley-Scott, N.~Bronn, K.~Naidoo, and E.~Reinert, ``Towards a semantic information theory (introducing quantum corollas),'' \emph{arXiv preprint arXiv:2201.05478}, Jan. 2022.

\bibitem{ChristoTWCArxiv2022}
C.~K. Thomas and W.~Saad, ``{Neuro-Symbolic Causal Reasoning Meets Signaling Game for Emergent Semantic Communications},'' \emph{IEEE Transactions on Wireless Communications}, Oct. 2023.

\bibitem{chaccour2022less}
C.~Chaccour, W.~Saad, M.~Debbah, Z.~Han, and H.~V. Poor, ``Less data, more knowledge: Building next generation semantic communication networks,'' \emph{arXiv preprint arXiv:2211.14343}, 2022.

\bibitem{chaccour2022disentangling}
C.~Chaccour and W.~Saad, ``Disentangling learnable and memorizable data via contrastive learning for semantic communications,'' in \emph{Proc. of the 56th Asilomar Conference on Signals, Systems, and Computers}, 2022, pp. 1175--1179.

\bibitem{schuld2019quantum}
M.~Schuld and N.~Killoran, ``Quantum machine learning in feature hilbert spaces,'' \emph{Physical review letters}, vol. 122, no.~4, Feb. 2019.

\bibitem{kerenidis2019q}
I.~Kerenidis, J.~Landman, A.~Luongo, and A.~Prakash, ``q-means: A quantum algorithm for unsupervised machine learning,'' \emph{Advances in Neural Information Processing Systems}, vol.~32, Dec. 2019.

\bibitem{cozzolino2019high}
D.~Cozzolino \emph{et~al.}, ``High-dimensional quantum communication: benefits, progress, and future challenges,'' \emph{Advanced Quantum Technologies}, vol.~2, no.~12, p. 1900038, Oct. 2019.

\bibitem{king2003capacity}
C.~King, ``The capacity of the quantum depolarizing channel,'' \emph{IEEE Transactions on Information Theory}, vol.~49, no.~1, pp. 221--229, Jan. 2003.

\bibitem{pan2001entanglement}
J.-W. Pan \emph{et~al.}, ``Entanglement purification for quantum communication,'' \emph{Nature}, vol. 410, no. 6832, pp. 1067--1070, April 2001.

\bibitem{StavrouTCOM2023}
P.~A. Stavrou and M.~Kountouris, ``{The role of fidelity in goal-oriented semantic communication: A rate distortion approach},'' \emph{IEEE Transactions on Communications}, vol.~71, Jul. 2023.

\bibitem{3gpp2007lte}
3GPP, ``\text{LTE} physical layer framework for performance verification,'' \emph{TS R1--070674, Third Generation Partnership Project (3GPP)}, Feb. 2007.

\bibitem{navarro2020survey}
Navarro-Ortiz \emph{et~al.}, ``A survey on {5G} usage scenarios and traffic models,'' \emph{IEEE Communications Surveys \& Tutorials}, vol.~22, no.~2, pp. 905--929, Feb. 2020.

\bibitem{WildeCUP2013}
M.~M. Wilde, ``Quantum information theory,'' in \emph{Cambridge university press}, 2013.

\end{thebibliography}
\end{spacing}

\vspace{-2mm}\appendices

\section{Proof of Lemma~\ref{theorem_se_rep}}
\label{appendix_proof_theorem_se_rep}
The average number of quantum communication resources conveyed by $\mathcal{S}$ will be $C$, which can be lower bounded by $C  \geq \sum\limits_{c_k\in \mathcal{C}}p( c_k) \mathbb{H}_q(\ket{y}\mid c_k)$ using the quantum analogue of Kraft's inequality \cite{WildeCUP2013}. Moreover, we can upper bound $C$ for the semantic representations by $\ln d_2$ using the results in \cite{WildeCUP2013}.

For a semantic-agnostic quantum system, where the semantic concepts are not extracted, and classical data is directly embedded using amplitude encoding, the lower bound becomes $C \geq \mathbb{H}_q(\ket{y})$,
and the upperbound becomes $\ln d_1$ \cite{WildeCUP2013}. 

The bounds for the QSC framework are lower as it only embeds quantum states carrying semantic concepts and compresses all quantum states that convey equivalent meanings from the receiver's standpoint. In contrast, semantic-agnostic QCNs employ data statistics of $\ket{y}$ for encoding and do not account for the semantics present in the data. This leads to a larger state space $\mathcal{Y}$ for semantic agnostic quantum systems as opposed to QSC, which employs a reduced state space $\mathcal{S}$. 

\end{document}